\documentclass[pra,aps,twocolumn,epsfig,showpacs,superscriptaddress,nofootinbib]{revtex4}
\usepackage{amsmath}
\usepackage{amssymb}
\usepackage[dvips]{graphicx}
\usepackage{dcolumn}
\usepackage{bm}
\usepackage[colorlinks=false,dvipdfm]{hyperref} 
\usepackage{setspace}
\usepackage{amsfonts}
\usepackage{rotating}
\usepackage{makeidx}
\usepackage{amsthm}

\begin{document}

\title{Sequential State Discrimination and Requirement of Quantum Dissonance\footnote{Physical Review A 88, 052331, (2013).}}

\author{Chao-Qian Pang }
\affiliation{Physics Department, School of Science, Tianjin
University, Tianjin 300072, China}

\author{Fu-Lin Zhang}
\email[Corresponding author: ]{flzhang@tju.edu.cn}
\affiliation{Physics Department, School of Science, Tianjin
University, Tianjin 300072, China}

\author{Li-Fang Xu}
\affiliation{Physics Department, School of Science, Tianjin
University, Tianjin 300072, China}

\author{Mai-Lin Liang}
\affiliation{Physics Department, School of Science, Tianjin
University, Tianjin 300072, China}

\author{Jing-Ling Chen }
\email[Corresponding author
: ]{chenjl@nankai.edu.cn}
\affiliation{Theoretical
Physics Division, Chern Institute of Mathematics, Nankai University,
Tianjin, 300071, China} \affiliation{Centre for Quantum
Technologies, National University of Singapore, 3 Science Drive 2,
Singapore 117543}

\date{\today}

\begin{abstract}
We study the procedure for sequential unambiguous state discrimination.
 A qubit is prepared in one of two possible states, and measured by two observers Bob and Charlie sequentially.
A necessary condition for the state to be unambiguously discriminated by Charlie is the absence of entanglement between the principal qubit, prepared by Alice, and Bob's auxiliary system.
In general, the procedure for both Bob and Charlie to recognize between two nonorthogonal states conclusively relies on the availability of quantum discord
 which is precisely the quantum dissonance when the entanglement is absent. In Bob's measurement, the left discord is positively correlated with the information extracted by Bob, and the right discord enhances the information left to Charlie.
When their product achieves its maximum the probability for both Bob and Charlie to identify the state achieves its optimal value.
\end{abstract}



\maketitle



\section{Introduction}
It is always interesting to uncover non-trivial roles of
quantum correlations contained in a composite quantum system in quantum information processing.
 These correlations originate from quantum coherent superposition and have been widely studied in various perspectives, such as quantum entanglement \cite{RevModPhys.81.865}, Bell nonlocality \cite{Bell}, quantum discord \cite{PhysRevLett.88.017901,henderson2001classical}, and so on.
Quantum entanglement was believed to be the only key resource in quantum information processing.
However, recent developments \cite{lanyon2008experimental,datta2008quantum} demonstrated that, the algorithm for deterministic quantum computation
with one qubit (DQC1)
can surpass the performance of its corresponding
classical algorithm in the absence of quantum entanglement between
the the control qubit and a completely mixed state.
The quantum discord is considered as a main reason for the advantage of
the process of DQC1 without entanglement, and thereby has gained wide attention in recent years \cite{modi2010unified,PhysRevA.85.032104,PhysRevA.86.012312}.
Particularly, a new measure of correlation in a bipartite quantum system called quantum dissonance has been introduced which can be expressed by a unified view of quantum correlations based on the relative entropy \cite{modi2010unified} and be regarded as the entanglement-excluded nonclassical correlation.
The quantum dissonance of a separable state (with zero entanglement) is
exactly equal to its discord. Therefore, the
quantum discord between the control qubit and the input state in DQC1, which plays a key role in the computational process, is nothing but the quantum dissonance.

The quantum algorithm in DQC1 is not the unique case where the quantum dissonance serves as a key resource.
Roa \emph{et. al.} \cite{roa2011dissonance} found that when two nonorthogonal states are prepared with equal \emph{a priori} probabilities, the quantum dissonance is the only quantum correlation, which is
required for performing optimal unambiguous quantum state discrimination with the assistance of an auxiliary
qubit.
Immediately afterwards, Zhang \emph{et. al.} \cite{zhang2012assisted} proved that the quantum entanglement in a more general protocol was completely unnecessary as it can be zero for the arbitrary \emph{a priori} probabilities.
These results indicate that the procedure for assisted optimal state discrimination (AOSD) is the second example after DQC1 that can be implemented successfully, aided only by quantum dissonance rather than entanglement.

Unambiguous discrimination among linearly independent nonorthogonal quantum states
 is a fundamental problem both in quantum mechanics and quantum information theory \cite{ivanovic1987differentiate,peres1988differentiate,dieks1988overlap,john1996mathematical,helstrom1969quantum,roa2002quantum,POVM2008,POVM2009,bennett,mimih,bhh}.
In its simplest form, the task of observer Bob is to determine, with no error permitted, what the state of the qubit is, which is prepared by Alice in one of two known states, $|\psi_{1}\rangle$ or $|\psi_{2}\rangle$.
Since $|\psi_{1}\rangle$ and $|\psi_{2}\rangle$ are not orthogonal, the distinction may sometimes fail as the the price to pay for no error.
Consequently, Bob's measurement has three possible outcomes, $|\psi_{1}\rangle$, $|\psi_{2}\rangle$, and inconclusive.

In their very recent work \cite{bergou2013extracting}, Bergou \emph{et. al.} developed a theory of \emph{nondestructive} sequential
quantum measurements based on the unambiguous quantum state discrimination.
It originated from the topic of extracting information from a quantum system by multiple observers \cite{nagali2012testing,PhysRevA.84.032326,PhysRevA.83.032311}.
 Namely, they added another observer Charlie who also performed an unambiguous discrimination measurement on the same qubit after Bob's measurement.
As a scenario typical in secure quantum communication strategies, Alice, Bob, and Charlie can do any
pre-measurement conspiracy, but no classical communication is allowed after Bob's measurement.
The probability for both Bob and Charlie to identify the state is found with a nonzero value.
The probability that Charlie's measurement succeeds, which quantifies the information about the state Alice sent is left in the qubit after Bob's measurement, depends on the overlap between the two possible states that Bob's measurement leaves in the qubit.

A two-state quantum system, or qubit, can be represented by a two-dimensional Hilbert space.
A three-dimensional Hilbert space is required to implement an optimal procedure of unambiguous discrimination between two nonorthogonal states\cite{roa2002quantum}.
The observers have to introduce ancillary systems to increase the
dimension of the Hilbert space \cite{roa2002quantum}.
This leads to a natural question: What kinds of correlations, entanglement, quantum discord,
or dissonance, serve as a key resource, especially for the information left by Bob for
Charlie to measure, in sequential state discrimination?
 The original results in \cite{bergou2013extracting} were derived in the positive-operator-valued measure (POVM) formalism.
 In this work, we present a realization of the protocol in \cite{bergou2013extracting} by using
 the language of system ancilla and study the roles of correlations in the procedure.


\section{Sequential state discrimination}

\begin{figure}
\includegraphics[width=8cm]{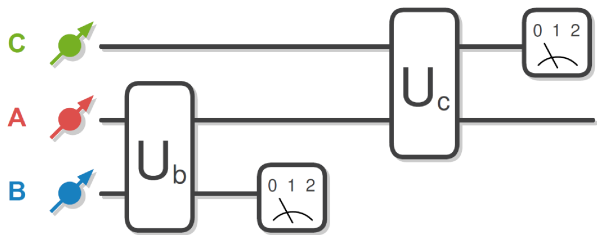} \\
\caption{(Color online) Protocol for sequential state discrimination. Alice prepares a qubit $A$ in one of two known states, $|\psi_{1}\rangle$ or $|\psi_{2}\rangle$ and sends it to Bob. Bob performs a joint unitary operation $\mathcal{U}_b$ between the qubit $A$ and his auxiliary qutrit $B$, followed by a von Neumann measurement
on the qutrit. His state discrimination is successful if the outcome of $B$ is 1 (for $|\psi_{1}\rangle$) and 2 (for $|\psi_{2}\rangle$), but unsuccessful if the outcome is 0. Then, Bob sends qubit $A$ in the post-measurement state to Charlie. Charlie makes a similar joint unitary $\mathcal{U}_c$ between it and his qutrit $C$, and performs a similar von Neumann measurement on $C$.
} \label{fig1}
\end{figure}

In Fig. \ref{fig1}, we show the procedure for sequential state discrimination.
Alice sends a qubit $A$ to Bob prepared in one of two nonorthogonal states, $|\psi_{1}\rangle$ or $|\psi_{2}\rangle$.
Bob has an auxiliary qutrit $B$ with the initial state $|0\rangle_b$. Performing a joint unitary
transformation  $\mathcal{U}_b$ between $A$ and $B$,
 Bob obtains the composite system in the states
\begin{subequations}\label{Utrans}
\begin{align}
\mathcal {U}_b\;|\psi_1\rangle|0\rangle_b=\sqrt{q_1^b}\;|\chi_1 \rangle|0\rangle_b+\sqrt{1-q_1^b}|\phi_1 \rangle|1\rangle_b, \\
\mathcal
{U}_b\;|\psi_2\rangle|0\rangle_b=\sqrt{q_2^b}\;|\chi_2 \rangle|0\rangle_b+\sqrt{1-q_2^b}|\phi_2 \rangle|2\rangle_b,
\end{align}
\end{subequations}
where $\{|0\rangle_b, |1\rangle_b, |2\rangle_b\}$ is the basis for the ancilla, and $|\chi \rangle_{1,2}$ and $|\phi \rangle_{1,2}$ are pure states
of $A$.
For the input state $|\psi_i \rangle$ ($i=1,2$), Bob's discrimination
is successful if his ancilla collapses to $| i \rangle_b$, and the system qubit $A$ collapses to $|\phi_i \rangle$ simultaneously. Otherwise, the process fails
when the projection is onto $| 0 \rangle_b$, and $A$ to $|\chi_i \rangle$.
The \emph{a priori} fixed overlap $\langle \psi_1 | \psi_2 \rangle $ does
not change due to the joint unitary transformation. Thus the
post-measurement states (failure states) $|\chi_i \rangle$ satisfy the constraint $\sqrt{q_1^b q_2^b}\langle \chi_1 | \chi_2 \rangle =\langle \psi_1 | \psi_2 \rangle$. Without loss of generality, we take the \emph{a priori} overlap $s=\langle\psi_{1}|\psi_{2}\rangle$ to be real ($0\leq s \leq 1$) in the present work.
As the simplest case, we assume that the two nonorthogonal states $|\psi_i \rangle$ ($i=1,2$) are prepared with equal \emph{a priori} probabilities.
Then, for a given value of $\langle \chi_1 | \chi_2 \rangle=t$,
the maximal success probability of Bob is attained for $q_1^b= q_2^b =s/t$, which is
\begin{equation}
P_{b}  = 1- s/t \ .
\label{Pb}
\end{equation}

After Bob's measurement, the qubit $A$ is sent to the second observer Charlie.
We assume that Charlie knows exactly what type of measurement Bob has performed, which is a necessary condition for Charlie to perform unambiguous discrimination.
In addition, the states sent to Charlie to be discriminated must be linearly independent.
Since the Hilbert space of a qubit is two dimensional, Charlie can only discriminate between two possible pure states.
This requires Bob's post-measurement states $|\chi_{i}\rangle= |\phi_{i}\rangle$.
Therefore, if Alice sent $|\psi_{i}\rangle$ ($i=1,2$), Charlie will receive $|\phi_{i}\rangle$, whether Bob's measurement succeeded or not.
Then, the states in Eq. (\ref{Utrans}) become
\begin{subequations}\label{Utrans1}
\begin{align}
\mathcal {U}_b\;|\psi_1\rangle|0\rangle_b= |\phi_1 \rangle|\eta_1\rangle_b, \\
\mathcal
{U}_b\;|\psi_2\rangle|0\rangle_b=|\phi_2 \rangle|\eta_2\rangle_b,
\end{align}
\end{subequations}
where $|\eta_i\rangle_b= \sqrt{q_i^b}  |0\rangle_b+\sqrt{1-q_i^b} |i\rangle_b$ with $i=1,2$.
 The unitary transformation has the form as
\begin{eqnarray}
\mathcal{U}_b=\frac{1}{1-s^2}\biggr(  |\phi_1 \rangle|\eta_1\rangle_b \langle \tilde{\psi}_1| _b\langle 0|
 + |\phi_2 \rangle|\eta_2\rangle_b \langle \tilde{\psi}_2| _b\langle 0|  \biggr) +
\mathcal{V}, \nonumber
\end{eqnarray}
where $|\tilde{\psi}_{i}\rangle= |\psi_{i}\rangle - |\psi_{j}\rangle
\langle \psi_{j} |\psi_{i}\rangle $ are the components of $
|\psi_{i}\rangle$ that orthogonal to $|\psi_{j}\rangle$ with $i \neq j$ and $i,j=1,2$,
and $\mathcal{V}$ denotes the terms operating on the subspace of $\{|1\rangle_b,|2\rangle_b \}$.
Here, one can obtain the POVM formalism in \cite{bergou2013extracting}.
The detection operators on the system qubit are $A_k^b= _b\langle k| \mathcal{U}_b |0\rangle_b$ ($k=0,1,2$),
and their corresponding POVM elements are given by  $\Pi_{k}^{b}=A_{k}^{b \dagger}A_{k}^b$ for $k=0,1,2$.

Charlie makes a similar joint unitary operation $\mathcal{U}_c$ between the qubit $A$ from Bob and his auxiliary qutrit $C$ as Eq. (\ref{Utrans}) with the parameters $q_1^c$ and $q_2^c$, followed by a von
Neumann measurement on $C$. His optimal success probability of unambiguous discrimination occurs when his two post-measurement states for the inconclusive outcomes are the same and $q_1^c = q_2^c =t$. The optimal success probability can be expressed as
\begin{equation}
P_{c}  = 1- t \ .
\label{Pc}
\end{equation}
It is negatively related to Bob's  success probability $P_b$ in Eq. (\ref{Pb}).
Its maximum $P_{c}  = 1- s $ corresponds to $P_b=0$, and minimum $P_{c}  = 0 $  corresponds to the maximal value of $P_b=1-s$.
These indicate that the quantum information carried by the qubit $A$ is finite,
and the overlap $t$ quantifies how much information about the state Alice sent is left in the qubit after
Bob's measurement.

The probability for both Bob and Charlie to identify the state is
\begin{equation}
P_{bc} = \frac{1}{2}[(1 - q_{1}^{c})(1 - q_{1}^{c}) + (1 - q_{2}^{b})(1 - q_{2}^{c})] .
\label{Pbc}
\end{equation}
The two unitary transformations $\mathcal{U}_{b,c}$ lead to two constraints $\sqrt{q_1^b q_2^b}  t =s $ and $\sqrt{q_1^c q_2^c} \leq t$.
The analysis of the optimal value of $P_{bc}$ can be divided into two cases:
(i) $s <  3-2 \sqrt{2}$, $P_{bc,\max}$ is attained for
$q_{1}^{b}=q_{2}^{b}=q_{1}^{c}=q_{2}^{c}=t=\sqrt{s}$; (ii) $s \geq  3-2 \sqrt{2}$, $P_{bc,\max}$ is attained for
$q_{1}^{b}=q_{1}^{c}=s$, $q_{2}^{c}=q_{2}^{b}=1$ and $t=\sqrt{s}$ (or exchange the subscripts $1$ and $2$). One has
\begin{subequations}\label{twocase}
\begin{align}
&&P_{bc,\max} =  (1-\sqrt{s} )^2, \ \ \ \ \ \ {s <  3-2 \sqrt{2}},\label{case1}&&\\
&&P_{bc,\max} = \frac{1}{2}(1-s)^2, \ \ \ \ \ \ {s \geq  3-2 \sqrt{2}}.\label{case2}&&
\end{align}
\end{subequations}
In case (i), the two states prepared by Alice are equally important in their protocol.
But the lack of quantum information in qubit $A$ leads to
a symmetry breaking in case (ii), where Bob and Charlie conspire to ignore one of the states \footnote{
Similar symmetry breaking also exists in the two strategies that allow Bob and
Charlie to communicate classically in the paper \cite{bergou2013extracting} by Bergou \emph{et. al.}.
There are two alternative optimal probabilities $ P_S^{(2)}= (1-s^2 )^2/2$ and $ P_S^{(3)}= (1-s^2 )^2/ 2 (1+s^2 ) $ beside the results in their Eqs. (15) and (16).}.


\section{quantum correlations}
In the procedure, the auxiliary systems $B$ and $C$ play the role of quantum detectors,
whose couplings with $A$ are key parts in the implementation of sequential state discrimination.
We now answer the question mentioned about what kind of quantum
correlation between $A$ and $B$ or $A$ and $C$ allows performing sequential state discrimination.
The discrimination probability of one observer is independent of the choice of the post-measurement states, which are corresponding to the successful outcomes as shown by Eqs. (\ref{Utrans}) and (\ref{Pb}).
Choosing $|\phi_i\rangle = |\chi_i\rangle$, Bob has the system-ancilla state
\begin{eqnarray}\label{rhoAB}
\rho_{AB} &=& \frac{1}{2} \mathcal {U}_b\left( |\psi_1\rangle\langle\psi_1|\otimes |0\rangle_b \langle 0| + |\psi_2\rangle \langle\psi_2|\otimes
|0\rangle_b \langle 0|\right) \mathcal {U}_b^\dag \nonumber \\
&=&  \frac{1}{2} \left( |\phi_1\rangle\langle\phi_1|\otimes |\eta_1\rangle_b \langle \eta_1| + |\phi_2\rangle \langle\phi_2|\otimes
|\eta_2\rangle_b \langle \eta_2|\right), \ \ \ \
\end{eqnarray}
which has a separable form obviously.
 Hence, the one-observer state discrimination process can be performed when the entanglement is absent.
In addition, the separable state (\ref{rhoAB}) precisely accords with the condition of Charlie's discrimination.
In other words, the entanglement between $A$ and $B$ is not only unnecessary for Bob's recognition, but also an obstacle
for the next observer Charlie.

The previous analysis prompts us to explore which correlation makes a positive effect on the procedure of sequential state discrimination.
The recent developments on the state discrimination assisted with an auxiliary qubit \cite{roa2011dissonance,zhang2012assisted,PhysRevA.85.022328}
inspired us to consider quantum dissonance as a candidate.
The dissonance of state (\ref{rhoAB}) is exactly equal to its discord while the entanglement is not allowed.

Quantum discord was presented based on the viewpoint that the total correlation in a bipartite system is divided
 into classical part and nonclassical one \cite{PhysRevLett.88.017901,henderson2001classical}.
The total correlation for a bipartite state $\rho_{AB}$ is defined
as its quantum mutual information $I(\rho_{AB})=S(\rho_A)+S(\rho_B)-S(\rho_{AB})$, where $S(\rho)$ is the von Neumann
entropy and $\rho_A$ and $\rho_B$ are two reduced states.
The classical correlation of the bipartite state is given by $J(\rho_{AB})=\sup_{ \{ E_k \} }\{S(\rho_A)-\sum_k p_{k} S(\rho_{A|k})\} $, where $p_{k}={\rm Tr}(\mathbb{I}_A\otimes E_k\rho_{AB})$ are the probabilities for outcomes $E_k$, $\rho_{A|k}$ are the partial projections $\rho_{A|k}={\rm Tr}_B(\mathbb{I}_A\otimes E_k\rho_{AB})/p_{k}$ , and the supreme is taken
over all the von Neumann measurement sets $\{E_k\}$ applied on subsystem $B$.
They are the quantum generalizations of two equivalent expressions for the classical
mutual information, and quantum discord is defined as the difference between them,
\begin{eqnarray}\label{discord}
D(\rho_{AB})=I(\rho_{AB})-J(\rho_{AB}).
\end{eqnarray}
It is a non-symmetrical quantity.
Similarly one can define the quantum discord $D(\rho_{BA})$
where one optimizes over measurement sets on subsystem $A$.
In this work, we call $D(\rho_{AB})$ as \emph{right discord} and $D(\rho_{BA})$ as \emph{left discord} according to \cite{dakic2010necessary}.

A zero-right-discord two-qubit state is of the form $\rho_{AB}=p_+ \rho_+ \otimes |\psi_+\rangle \langle \psi_+|+p_- \rho_- \otimes |\psi_-\rangle \langle \psi_-|$
,where $\rho_{\pm}$ are two density matrices for subsystem $A$, $p_{\pm}$
are non-negative numbers such that $p_+ + p_- =1$ and the two pure states of $B$ satisfy $|\langle \psi_+|\psi_- \rangle |=1$ or $0$ \cite{dakic2010necessary}.
The same conditions also exist for the left discord.
In state (\ref{rhoAB}), the overlap $\langle \eta_1 |\eta_2\rangle=s/t$ corresponds to the successful probability of  Bob's measurement,
 and $\langle \phi_1 |\phi_2\rangle=t$ for the state discrimination of the next observer Charlie.
One can easily find that the two discords are zero simultaneously when $\langle \eta_1 |\eta_2\rangle=0$ or $\langle \phi_1 |\phi_2\rangle=0$.
These suggest that the information extracted by Bob or Charlie from the qubit $A$ is influenced by both the left and the right discords.

\begin{figure}
\includegraphics[width=4.2cm]{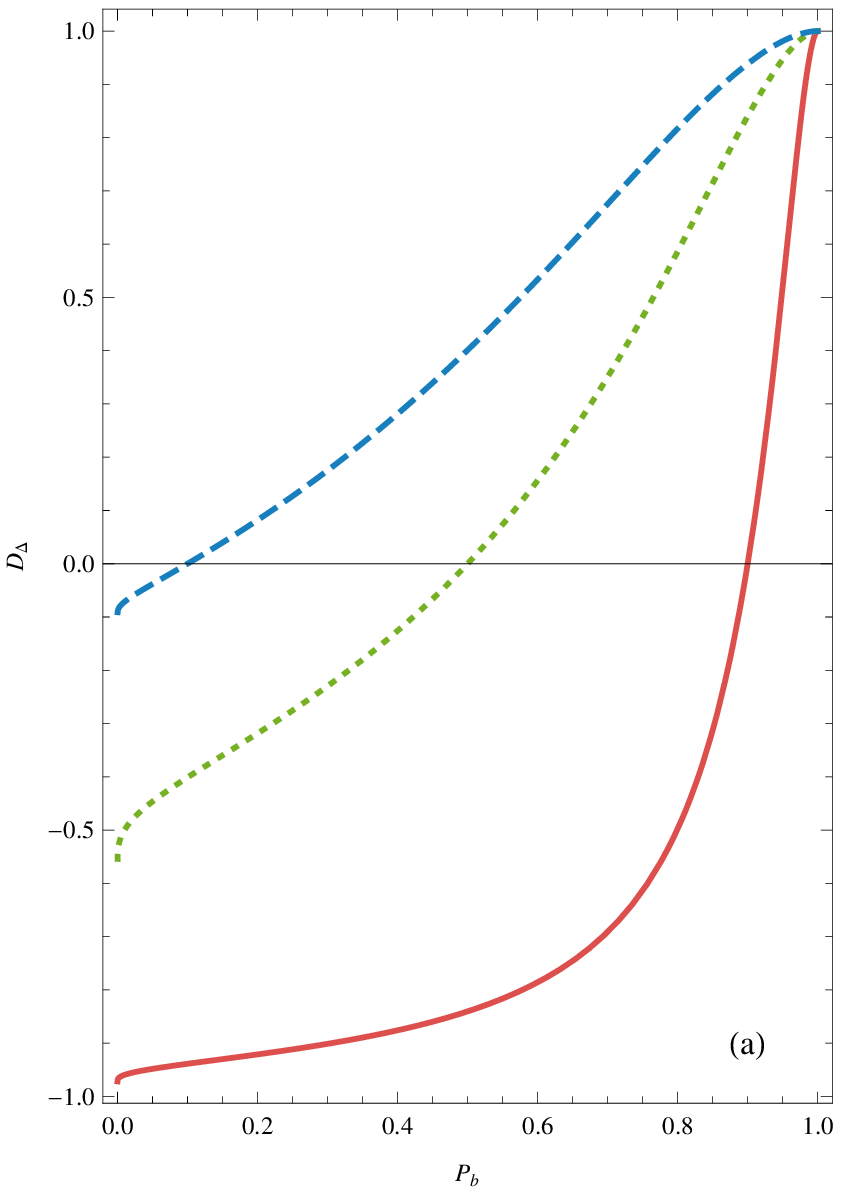} \
\includegraphics[width=4.2cm]{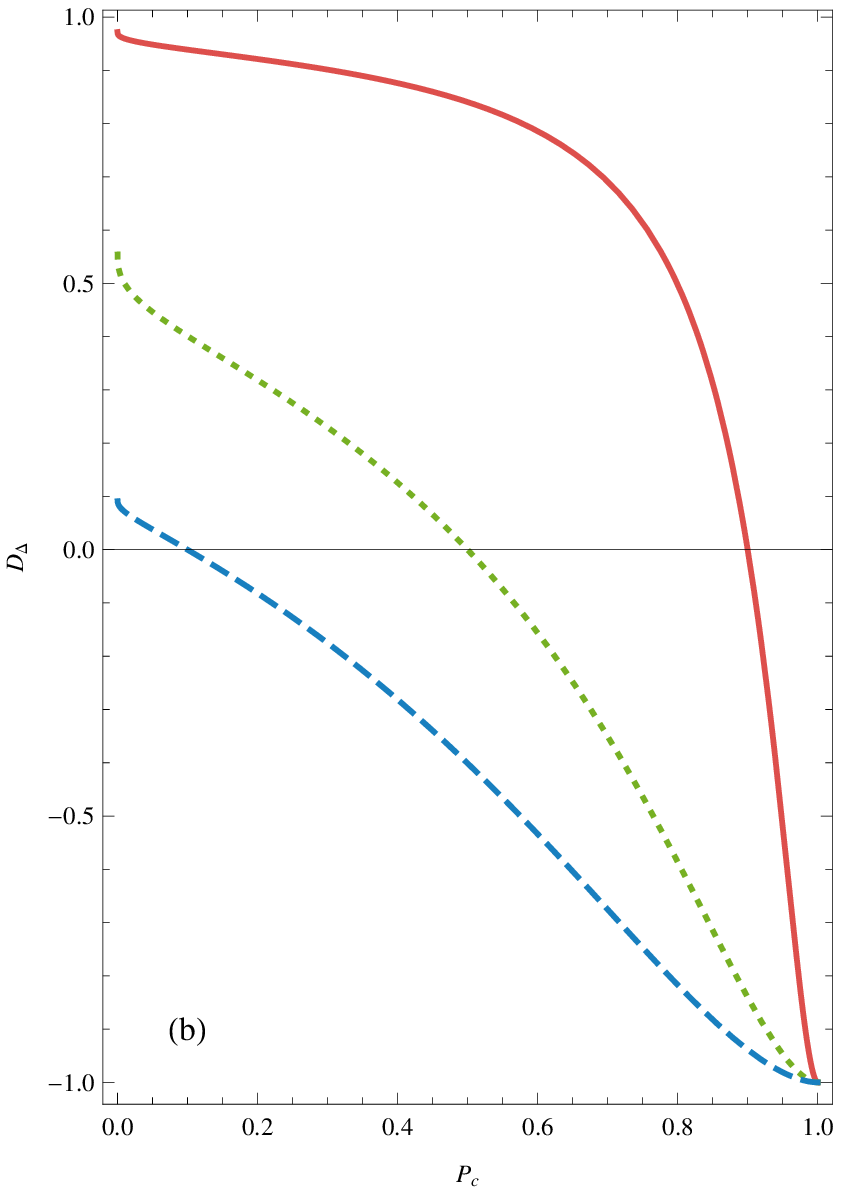} \\
\caption{(Color online) Relative difference $D_{\Delta}$ between two discords vs. success probability (a): $P_b$ or (b): $P_c$ with the other fixed. Solid lines: $P_{c \ \rm{or} \  b}=0.9$; dotted lines $P_{c \   \rm{or} \   b}=0.5$; dashed lines: $P_{c \  \rm{or} \   b}=0.1$.
} \label{fig2}
\end{figure}

\begin{figure}
\includegraphics[width=8cm]{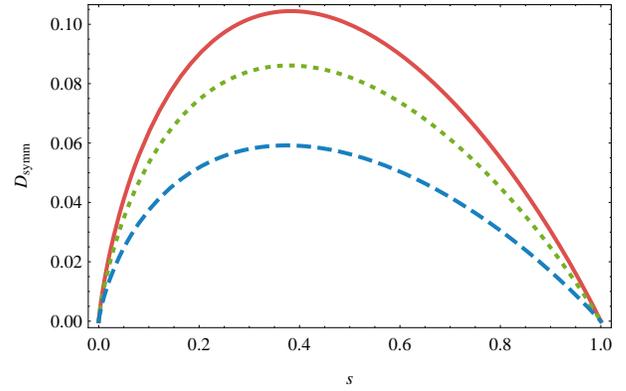}  \\
\caption{(Color online) The symmetrized discord in the sequential state discrimination procedure as a function of the overlap $s$, for $t=s^{1/2}$ (solid line), $t=s^{1/4}$  (dotted line), and $t=s^{1/8}$ (dashed line).
} \label{fig3}
\end{figure}

Since the state $\rho_{AB}$ has rank two, its discords can be derived analytically by using
the Koashi-Winter identity \cite{koashi2004monogamy}. It can be written as a reduced state of the tripartite pure state
\begin{eqnarray}\label{3qubit}
|\Psi \rangle =\frac{1}{\sqrt{2}}\bigr(|\phi_1\rangle|
\eta_1\rangle_b|0\rangle_d + |\phi_2\rangle|
\eta_2\rangle_b|1\rangle_d  \bigr),
\end{eqnarray}
where we have introduced a fictitious qubit $D$ with basis $\{|0\rangle_d, |1\rangle_d \}$.
The right discord $D(\rho_{AB})=S(\rho_B)-S(\rho_{AB})+E(\rho_{AD})=S(\rho_B)-S(\rho_{D})+E(\rho_{AD})$
where $E(\rho_{AD})$ is the
entanglement of formation \cite{Wootters98} between the principal
system $A$ and the qubit $D$.
Its explicit expression is
\begin{eqnarray}\label{discord1}
D(\rho_{AB})=\mathcal{H}(\tau_{B})-\mathcal{H}(\tau_{D})+\mathcal{H}(\tau_{A}-\tau_{ABD}),
\end{eqnarray}
where $\mathcal{H}(x)=-\frac{1+\sqrt{1-x}}{2}\log_{2}
\frac{1+\sqrt{1-x}}{2}-\frac{1-\sqrt{1-x}}{2}\log_{2}
\frac{1-\sqrt{1-x}}{2}$,
$\tau_{ABD}$ is the residual tangle (three-tangle) \cite{coffman2000distributed} of the tripartite state (\ref{3qubit}), and
$\tau_A$ (similarly for $\tau_B$ and $\tau_D$) is the tangle between $A$ and $BD$.
They are given by
\begin{eqnarray}
&&\tau_{ABD} = (1 - t^2)(1 - r^2), \nonumber \\
&&\tau_{A} =  1 - t^2 , \nonumber \\   &&\tau_{B} = 1 - r^2 ,\nonumber \\   &&\tau_{D} =  1 -  t^2r^2 ,
\end{eqnarray}
where we have set $r=s/t=\langle \eta_1 |\eta_2\rangle$. The left discord $D(\rho_{BA})$ can be easily obtained
by interchanging  $r$ and $t$ in $D(\rho_{AB})$.
 To analyze their roles in the protocol of sequential state discrimination, we define the relative difference between the two discords
\begin{eqnarray}
 D_{\Delta}=\frac{D(\rho_{BA})-D(\rho_{AB})}{D(\rho_{BA})+D(\rho_{AB})},
\end{eqnarray}
and a symmetrized discord
\begin{eqnarray}\label{Dsymm}
 D_{\textrm{symm}}=\sqrt{D(\rho_{BA})D(\rho_{AB})}.
\end{eqnarray}
In Fig. \ref{fig2}, we show the relation between the relative difference and the success probability $P_b$ or $P_c$ when the other is fixed.
For a given amount of $P_c$, when $P_b$ approximates $0$, the left discord $D(\rho_{BA})$ is less than the right one $ D(\rho_{AB})$.
 As $P_b$ increases, the relative difference $D_{\Delta}$ grows accordingly.
 When $P_b \rightarrow 1$, only the left discord exists in the state $\rho_{AB}$.
Hence, the information extracted by Bob from the qubit $A$ is enhanced by the left discord of $\rho_{AB}$, but suppressed by the right one.
The same analysis indicates that the right discord is positively correlated with the information left about the state Alice sent after
Bob's measurement, and the left discord has the opposite effect.
We also find that, for a fixed \emph{a priori} overlap $s=\langle \psi_1 | \psi_2 \rangle$, the symmetrized discord $D_{\textrm{symm}}$ achieves its maximum when $t=\sqrt{s}$, where occurs the optimal probability for both Bob and Charlie to identify the state. Also, as shown in Fig. \ref{fig3}, for a given power function relation between $t$ and $s$, there are
two cases for which a classical state appears: (i) two
orthogonal states for which the discrimination procedure becomes a von
Neumann measurement and (ii) two parallel states for which the information sent by Alice is zero.
Consequently, the symmetrized discord in state (\ref{rhoAB}) depends on the following factors: (a) the quantum information sent by Alice, (b) the joint probability for discriminations of Bob and Charlie, and (c) the deviation of their discrimination procedures from a von Neumann measurement.

\section{Summary}

We study the sequential unambiguous state discrimination by using
 the language of system ancilla and explore the roles of quantum correlations in the procedure.
We find that in the unambiguous discrimination of a single observer the entanglement between the principal qubit and the
ancilla is unnecessary.
In addition, the absence of entanglement is a necessary condition for the state to be unambiguously discriminated by the next observer.
  In general, the procedure for both Bob
and Charlie to recognize between two nonorthogonal states conclusively relies on the availability of
quantum discord which is precisely the quantum dissonance when the entanglement is absent.

The left discord in Bob's measurement enhances the information extracted by
Bob and suppresses the information left to Charlie.
 The role of the right discord is just the opposite.
The symmetrized discord defined as the square root of their product in Eq. (\ref{Dsymm}) is positively correlated
with the factors: the joint probability for discriminations of Bob and Charlie, the quantum information sent by Alice, and the deviation of their discrimination procedures from a von Neumann measurement.


\begin{acknowledgments}
F.L.Z. is supported by the NSF of China (Grant No. 11105097). J.L.C. is
supported by National Basic Research Program (973 Program) of China
under Grant No. 2012CB921900, NSF of China (Grants No. 10975075 and No.
11175089) and also partly supported by the National Research Foundation
and Ministry of Education, Singapore.
\end{acknowledgments}

\bibliography{MultDiscDiscord}

\begin{thebibliography}{31}
\expandafter\ifx\csname natexlab\endcsname\relax\def\natexlab#1{#1}\fi
\expandafter\ifx\csname bibnamefont\endcsname\relax
  \def\bibnamefont#1{#1}\fi
\expandafter\ifx\csname bibfnamefont\endcsname\relax
  \def\bibfnamefont#1{#1}\fi
\expandafter\ifx\csname citenamefont\endcsname\relax
  \def\citenamefont#1{#1}\fi
\expandafter\ifx\csname url\endcsname\relax
  \def\url#1{\texttt{#1}}\fi
\expandafter\ifx\csname urlprefix\endcsname\relax\def\urlprefix{URL }\fi
\providecommand{\bibinfo}[2]{#2}
\providecommand{\eprint}[2][]{\url{#2}}

\bibitem[{\citenamefont{Horodecki et~al.}(2009)\citenamefont{Horodecki,
  Horodecki, Horodecki, and Horodecki}}]{RevModPhys.81.865}
\bibinfo{author}{\bibfnamefont{R.}~\bibnamefont{Horodecki}},
  \bibinfo{author}{\bibfnamefont{P.}~\bibnamefont{Horodecki}},
  \bibinfo{author}{\bibfnamefont{M.}~\bibnamefont{Horodecki}},
  \bibnamefont{and}
  \bibinfo{author}{\bibfnamefont{K.}~\bibnamefont{Horodecki}},
  \bibinfo{journal}{Rev. Mod. Phys.} \textbf{\bibinfo{volume}{81}},
  \bibinfo{pages}{865} (\bibinfo{year}{2009}).

\bibitem[{\citenamefont{Bell}(1964)}]{Bell}
\bibinfo{author}{\bibfnamefont{J.~S.} \bibnamefont{Bell}},
  \bibinfo{journal}{Physics} \textbf{\bibinfo{volume}{1}}, \bibinfo{pages}{195}
  (\bibinfo{year}{1964}).

\bibitem[{\citenamefont{Ollivier and Zurek}(2001)}]{PhysRevLett.88.017901}
\bibinfo{author}{\bibfnamefont{H.}~\bibnamefont{Ollivier}} \bibnamefont{and}
  \bibinfo{author}{\bibfnamefont{W.~H.} \bibnamefont{Zurek}},
  \bibinfo{journal}{Phys. Rev. Lett.} \textbf{\bibinfo{volume}{88}},
  \bibinfo{pages}{017901} (\bibinfo{year}{2001}).

\bibitem[{\citenamefont{Henderson and Vedral}(2001)}]{henderson2001classical}
\bibinfo{author}{\bibfnamefont{L.}~\bibnamefont{Henderson}} \bibnamefont{and}
  \bibinfo{author}{\bibfnamefont{V.}~\bibnamefont{Vedral}},
  \bibinfo{journal}{J. Phys. A} \textbf{\bibinfo{volume}{34}},
  \bibinfo{pages}{6899} (\bibinfo{year}{2001}).

\bibitem[{\citenamefont{Lanyon et~al.}(2008)\citenamefont{Lanyon, Barbieri,
  Almeida, and White}}]{lanyon2008experimental}
\bibinfo{author}{\bibfnamefont{B.~P.} \bibnamefont{Lanyon}},
  \bibinfo{author}{\bibfnamefont{M.}~\bibnamefont{Barbieri}},
  \bibinfo{author}{\bibfnamefont{M.~P.} \bibnamefont{Almeida}},
  \bibnamefont{and} \bibinfo{author}{\bibfnamefont{A.~G.} \bibnamefont{White}},
  \bibinfo{journal}{Phys. Rev. Lett.} \textbf{\bibinfo{volume}{101}},
  \bibinfo{pages}{200501} (\bibinfo{year}{2008}).

\bibitem[{\citenamefont{Datta et~al.}(2008)\citenamefont{Datta, Shaji, and
  Caves}}]{datta2008quantum}
\bibinfo{author}{\bibfnamefont{A.}~\bibnamefont{Datta}},
  \bibinfo{author}{\bibfnamefont{A.}~\bibnamefont{Shaji}}, \bibnamefont{and}
  \bibinfo{author}{\bibfnamefont{C.~M.} \bibnamefont{Caves}},
  \bibinfo{journal}{Phys. Rev. Lett.} \textbf{\bibinfo{volume}{100}},
  \bibinfo{pages}{050502} (\bibinfo{year}{2008}).

\bibitem[{\citenamefont{Modi et~al.}(2010)\citenamefont{Modi, Paterek, Son,
  Vedral, and Williamson}}]{modi2010unified}
\bibinfo{author}{\bibfnamefont{K.}~\bibnamefont{Modi}},
  \bibinfo{author}{\bibfnamefont{T.}~\bibnamefont{Paterek}},
  \bibinfo{author}{\bibfnamefont{W.}~\bibnamefont{Son}},
  \bibinfo{author}{\bibfnamefont{V.}~\bibnamefont{Vedral}}, \bibnamefont{and}
  \bibinfo{author}{\bibfnamefont{M.}~\bibnamefont{Williamson}},
  \bibinfo{journal}{Phys. Rev. Lett.} \textbf{\bibinfo{volume}{104}},
  \bibinfo{pages}{080501} (\bibinfo{year}{2010}).

\bibitem[{\citenamefont{Bellomo
  et~al.}(2012{\natexlab{a}})\citenamefont{Bellomo, Giorgi, Galve, Lo~Franco,
  Compagno, and Zambrini}}]{PhysRevA.85.032104}
\bibinfo{author}{\bibfnamefont{B.}~\bibnamefont{Bellomo}},
  \bibinfo{author}{\bibfnamefont{G.~L.} \bibnamefont{Giorgi}},
  \bibinfo{author}{\bibfnamefont{F.}~\bibnamefont{Galve}},
  \bibinfo{author}{\bibfnamefont{R.}~\bibnamefont{Lo~Franco}},
  \bibinfo{author}{\bibfnamefont{G.}~\bibnamefont{Compagno}}, \bibnamefont{and}
  \bibinfo{author}{\bibfnamefont{R.}~\bibnamefont{Zambrini}},
  \bibinfo{journal}{Phys. Rev. A} \textbf{\bibinfo{volume}{85}},
  \bibinfo{pages}{032104} (\bibinfo{year}{2012}{\natexlab{a}}).

\bibitem[{\citenamefont{Bellomo
  et~al.}(2012{\natexlab{b}})\citenamefont{Bellomo, Lo~Franco, and
  Compagno}}]{PhysRevA.86.012312}
\bibinfo{author}{\bibfnamefont{B.}~\bibnamefont{Bellomo}},
  \bibinfo{author}{\bibfnamefont{R.}~\bibnamefont{Lo~Franco}},
  \bibnamefont{and} \bibinfo{author}{\bibfnamefont{G.}~\bibnamefont{Compagno}},
  \bibinfo{journal}{Phys. Rev. A} \textbf{\bibinfo{volume}{86}},
  \bibinfo{pages}{012312} (\bibinfo{year}{2012}{\natexlab{b}}).

\bibitem[{\citenamefont{Roa et~al.}(2011)\citenamefont{Roa, Retamal, and
  Alid-Vaccarezza}}]{roa2011dissonance}
\bibinfo{author}{\bibfnamefont{L.}~\bibnamefont{Roa}},
  \bibinfo{author}{\bibfnamefont{J.~C.} \bibnamefont{Retamal}},
  \bibnamefont{and}
  \bibinfo{author}{\bibfnamefont{M.}~\bibnamefont{Alid-Vaccarezza}},
  \bibinfo{journal}{Phys. Rev. Lett.} \textbf{\bibinfo{volume}{107}},
  \bibinfo{pages}{080401} (\bibinfo{year}{2011}).

\bibitem[{\citenamefont{Zhang et~al.}(2013)\citenamefont{Zhang, Chen, Kwek, and
  Vedral}}]{zhang2012assisted}
\bibinfo{author}{\bibfnamefont{F.-L.} \bibnamefont{Zhang}},
  \bibinfo{author}{\bibfnamefont{J.-L.} \bibnamefont{Chen}},
  \bibinfo{author}{\bibfnamefont{L.~C.} \bibnamefont{Kwek}}, \bibnamefont{and}
  \bibinfo{author}{\bibfnamefont{V.}~\bibnamefont{Vedral}},
  \bibinfo{journal}{Sci. Rep.} \textbf{\bibinfo{volume}{3}},
  \bibinfo{pages}{2134} (\bibinfo{year}{2013}).

\bibitem[{\citenamefont{Ivanovic}(1987)}]{ivanovic1987differentiate}
\bibinfo{author}{\bibfnamefont{I.~D.} \bibnamefont{Ivanovic}},
  \bibinfo{journal}{Phys. Lett. A} \textbf{\bibinfo{volume}{123}},
  \bibinfo{pages}{257} (\bibinfo{year}{1987}).

\bibitem[{\citenamefont{Peres}(1988)}]{peres1988differentiate}
\bibinfo{author}{\bibfnamefont{A.}~\bibnamefont{Peres}},
  \bibinfo{journal}{Phys. Lett. A} \textbf{\bibinfo{volume}{128}},
  \bibinfo{pages}{19} (\bibinfo{year}{1988}).

\bibitem[{\citenamefont{Dieks}(1988)}]{dieks1988overlap}
\bibinfo{author}{\bibfnamefont{D.}~\bibnamefont{Dieks}},
  \bibinfo{journal}{Phys. Lett. A} \textbf{\bibinfo{volume}{126}},
  \bibinfo{pages}{303} (\bibinfo{year}{1988}).

\bibitem[{\citenamefont{Neumann}(1996)}]{john1996mathematical}
\bibinfo{author}{\bibfnamefont{J.~V.} \bibnamefont{Neumann}},
  \emph{\bibinfo{title}{Mathematical Foundations of Quantum Mechanics}},
  vol.~\bibinfo{volume}{2} (\bibinfo{publisher}{Princeton University Press,
  Princeton, NJ}, \bibinfo{year}{1996}).

\bibitem[{\citenamefont{Helstrom}(1969)}]{helstrom1969quantum}
\bibinfo{author}{\bibfnamefont{C.}~\bibnamefont{Helstrom}},
  \bibinfo{journal}{J. Stat. Phys.} \textbf{\bibinfo{volume}{1}},
  \bibinfo{pages}{231} (\bibinfo{year}{1969}).

\bibitem[{\citenamefont{Roa et~al.}(2002)\citenamefont{Roa, Retamal, and
  Saavedra}}]{roa2002quantum}
\bibinfo{author}{\bibfnamefont{L.}~\bibnamefont{Roa}},
  \bibinfo{author}{\bibfnamefont{J.}~\bibnamefont{Retamal}}, \bibnamefont{and}
  \bibinfo{author}{\bibfnamefont{C.}~\bibnamefont{Saavedra}},
  \bibinfo{journal}{Phys. Rev. A} \textbf{\bibinfo{volume}{66}},
  \bibinfo{pages}{012103} (\bibinfo{year}{2002}).

\bibitem[{\citenamefont{Jafarizadeh et~al.}(2008)\citenamefont{Jafarizadeh,
  Rezaei, Karimi, and Amiri}}]{POVM2008}
\bibinfo{author}{\bibfnamefont{M.~A.} \bibnamefont{Jafarizadeh}},
  \bibinfo{author}{\bibfnamefont{M.}~\bibnamefont{Rezaei}},
  \bibinfo{author}{\bibfnamefont{N.}~\bibnamefont{Karimi}}, \bibnamefont{and}
  \bibinfo{author}{\bibfnamefont{A.~R.} \bibnamefont{Amiri}},
  \bibinfo{journal}{Phys. Rev. A} \textbf{\bibinfo{volume}{77}},
  \bibinfo{pages}{042314} (\bibinfo{year}{2008}).

\bibitem[{\citenamefont{Pang and Wu}(2009)}]{POVM2009}
\bibinfo{author}{\bibfnamefont{S.}~\bibnamefont{Pang}} \bibnamefont{and}
  \bibinfo{author}{\bibfnamefont{S.}~\bibnamefont{Wu}}, \bibinfo{journal}{Phys.
  Rev. A} \textbf{\bibinfo{volume}{80}}, \bibinfo{pages}{052320}
  (\bibinfo{year}{2009}).

\bibitem[{\citenamefont{Bennett}(1992)}]{bennett}
\bibinfo{author}{\bibfnamefont{C.~H.} \bibnamefont{Bennett}},
  \bibinfo{journal}{Phys. Rev. Lett.} \textbf{\bibinfo{volume}{68}},
  \bibinfo{pages}{3121} (\bibinfo{year}{1992}).

\bibitem[{\citenamefont{Hillery and Mimih}(2003)}]{mimih}
\bibinfo{author}{\bibfnamefont{M.}~\bibnamefont{Hillery}} \bibnamefont{and}
  \bibinfo{author}{\bibfnamefont{J.}~\bibnamefont{Mimih}},
  \bibinfo{journal}{Phys. Rev. A} \textbf{\bibinfo{volume}{67}},
  \bibinfo{pages}{042304} (\bibinfo{year}{2003}).

\bibitem[{\citenamefont{Bergou et~al.}(2003)\citenamefont{Bergou, Herzog, and
  Hillery}}]{bhh}
\bibinfo{author}{\bibfnamefont{J.~A.} \bibnamefont{Bergou}},
  \bibinfo{author}{\bibfnamefont{U.}~\bibnamefont{Herzog}}, \bibnamefont{and}
  \bibinfo{author}{\bibfnamefont{M.}~\bibnamefont{Hillery}},
  \bibinfo{journal}{Phys. Rev. Lett.} \textbf{\bibinfo{volume}{90}},
  \bibinfo{pages}{257901} (\bibinfo{year}{2003}).

\bibitem[{\citenamefont{Bergou et~al.}(2013)\citenamefont{Bergou, Feldman, and
  Hillery}}]{bergou2013extracting}
\bibinfo{author}{\bibfnamefont{J.}~\bibnamefont{Bergou}},
  \bibinfo{author}{\bibfnamefont{E.}~\bibnamefont{Feldman}}, \bibnamefont{and}
  \bibinfo{author}{\bibfnamefont{M.}~\bibnamefont{Hillery}},
  \bibinfo{journal}{Phys. Rev. Lett.} \textbf{\bibinfo{volume}{111}},
  \bibinfo{pages}{100501} (\bibinfo{year}{2013}).

\bibitem[{\citenamefont{Nagali et~al.}(2012)\citenamefont{Nagali, Felicetti,
  de~Assis, D'Ambrosio, Filip, and Sciarrino}}]{nagali2012testing}
\bibinfo{author}{\bibfnamefont{E.}~\bibnamefont{Nagali}},
  \bibinfo{author}{\bibfnamefont{S.}~\bibnamefont{Felicetti}},
  \bibinfo{author}{\bibfnamefont{P.-L.} \bibnamefont{de~Assis}},
  \bibinfo{author}{\bibfnamefont{V.}~\bibnamefont{D'Ambrosio}},
  \bibinfo{author}{\bibfnamefont{R.}~\bibnamefont{Filip}}, \bibnamefont{and}
  \bibinfo{author}{\bibfnamefont{F.}~\bibnamefont{Sciarrino}},
  \bibinfo{journal}{Sci. Rep.} \textbf{\bibinfo{volume}{2}},
  \bibinfo{pages}{443} (\bibinfo{year}{2012}).

\bibitem[{\citenamefont{Rap\ifmmode~\check{c}\else \v{c}\fi{}an
  et~al.}(2011)\citenamefont{Rap\ifmmode~\check{c}\else \v{c}\fi{}an,
  Calsamiglia, Mu\~noz Tapia, Bagan, and Bu\ifmmode~\check{z}\else
  \v{z}\fi{}ek}}]{PhysRevA.84.032326}
\bibinfo{author}{\bibfnamefont{P.}~\bibnamefont{Rap\ifmmode~\check{c}\else
  \v{c}\fi{}an}},
  \bibinfo{author}{\bibfnamefont{J.}~\bibnamefont{Calsamiglia}},
  \bibinfo{author}{\bibfnamefont{R.}~\bibnamefont{Mu\~noz Tapia}},
  \bibinfo{author}{\bibfnamefont{E.}~\bibnamefont{Bagan}}, \bibnamefont{and}
  \bibinfo{author}{\bibfnamefont{V.}~\bibnamefont{Bu\ifmmode~\check{z}\else
  \v{z}\fi{}ek}}, \bibinfo{journal}{Phys. Rev. A}
  \textbf{\bibinfo{volume}{84}}, \bibinfo{pages}{032326}
  (\bibinfo{year}{2011}).

\bibitem[{\citenamefont{Filip}(2011)}]{PhysRevA.83.032311}
\bibinfo{author}{\bibfnamefont{R.}~\bibnamefont{Filip}},
  \bibinfo{journal}{Phys. Rev. A} \textbf{\bibinfo{volume}{83}},
  \bibinfo{pages}{032311} (\bibinfo{year}{2011}).

\bibitem[{\citenamefont{Li et~al.}(2012)\citenamefont{Li, Fei, Wang, and
  Fan}}]{PhysRevA.85.022328}
\bibinfo{author}{\bibfnamefont{B.}~\bibnamefont{Li}},
  \bibinfo{author}{\bibfnamefont{S.-M.} \bibnamefont{Fei}},
  \bibinfo{author}{\bibfnamefont{Z.-X.} \bibnamefont{Wang}}, \bibnamefont{and}
  \bibinfo{author}{\bibfnamefont{H.}~\bibnamefont{Fan}},
  \bibinfo{journal}{Phys. Rev. A} \textbf{\bibinfo{volume}{85}},
  \bibinfo{pages}{022328} (\bibinfo{year}{2012}).

\bibitem[{\citenamefont{Daki{\'c} et~al.}(2010)\citenamefont{Daki{\'c}, Vedral,
  and Brukner}}]{dakic2010necessary}
\bibinfo{author}{\bibfnamefont{B.}~\bibnamefont{Daki{\'c}}},
  \bibinfo{author}{\bibfnamefont{V.}~\bibnamefont{Vedral}}, \bibnamefont{and}
  \bibinfo{author}{\bibfnamefont{{\v{C}}.}~\bibnamefont{Brukner}},
  \bibinfo{journal}{Phys. Rev. Lett.} \textbf{\bibinfo{volume}{105}},
  \bibinfo{pages}{190502} (\bibinfo{year}{2010}).

\bibitem[{\citenamefont{Koashi and Winter}(2004)}]{koashi2004monogamy}
\bibinfo{author}{\bibfnamefont{M.}~\bibnamefont{Koashi}} \bibnamefont{and}
  \bibinfo{author}{\bibfnamefont{A.}~\bibnamefont{Winter}},
  \bibinfo{journal}{Phys. Rev. A} \textbf{\bibinfo{volume}{69}},
  \bibinfo{pages}{022309} (\bibinfo{year}{2004}).

\bibitem[{\citenamefont{Wootters}(1998)}]{Wootters98}
\bibinfo{author}{\bibfnamefont{W.~K.} \bibnamefont{Wootters}},
  \bibinfo{journal}{Phys. Rev. Lett.} \textbf{\bibinfo{volume}{80}},
  \bibinfo{pages}{2245} (\bibinfo{year}{1998}).

\bibitem[{\citenamefont{Coffman et~al.}(2000)\citenamefont{Coffman, Kundu, and
  Wootters}}]{coffman2000distributed}
\bibinfo{author}{\bibfnamefont{V.}~\bibnamefont{Coffman}},
  \bibinfo{author}{\bibfnamefont{J.}~\bibnamefont{Kundu}}, \bibnamefont{and}
  \bibinfo{author}{\bibfnamefont{W.~K.} \bibnamefont{Wootters}},
  \bibinfo{journal}{Phys. Rev. A} \textbf{\bibinfo{volume}{61}},
  \bibinfo{pages}{052306} (\bibinfo{year}{2000}).

\end{thebibliography}

\end{document}